\documentclass[letterpaper, 10pt, conference]{ieeeconf} 

\usepackage{amsmath}

\usepackage{array}
\usepackage{cite}
\usepackage{amsfonts}
\usepackage{amssymb}
\usepackage{graphicx} 
\usepackage{epstopdf}
\usepackage{float}
\usepackage{algorithm}
\usepackage{algorithmicx}
\usepackage{algpseudocode}
\usepackage{balance}
\usepackage[fancythm,fancybb]{jphmacros2e} 

\usepackage{amsfonts}
\usepackage{mathrsfs}

\usepackage[norefs,nocites]{refcheck}
\usepackage{relsize}
\usepackage{capt-of}
\usepackage{mathtools}

\IEEEoverridecommandlockouts                              
\algdef{SE}[FOR]{NoDoFor}{EndFor}[1]{\algorithmicfor\ #1}{\algorithmicend\ \algorithmicfor}
\algdef{SE}[IF]{NoThenIf}{EndIf}[1]{\algorithmicif\ #1}{\algorithmicend\ \algorithmicif}%
\algdef{SE}[WHILE]{NoDoWhile}{EndWhile}[1]{\algorithmicwhile\ #1}{\algorithmicend\ \algorithmicwhile}%

\def\frqed{\ifhmode\nobreak\hbox to5pt{\hfil}\nobreak%
	\hskip 0pt plus1fill\nobreak\fi\quad\qedsymbol\renewcommand{\qed}{}} 

\newcommand\norm[1]{\left\lVert#1\right\rVert}
\IEEEoverridecommandlockouts                              

\allowdisplaybreaks
\begin{document}

\newtheorem{theorem}{Theorem}
\newtheorem{fact}{Fact}
\newtheorem{lemma}{Lemma}

\title{A Predictive Deep Learning Approach to Output Regulation: The Case of Collaborative Pursuit Evasion}
\author{S.~Shivam$^1$,
~A.~Kanellopoulos$^2$, 
        ~K.~G.~Vamvoudakis$^2$,
   and      Y.~Wardi$^1$ 
\thanks{$^1$S.~Shivam, and Y.~Wardi are with the School of Electrical and Computer Engineering, Georgia Institute of Technology, Atlanta, GA, $30332$, USA.  e-mail: (sshivam6@gatech.edu, ywardi@gatech.edu).}
\thanks{$^2$A.~Kanellopoulos, and K.~G.~Vamvoudakis are with the Daniel Guggenheim School of Aerospace Engineering, Georgia Institute of Technology, Atlanta, GA, $30332$, USA. e-mail: (ariskan@gatech.edu, kyriakos@gatech.edu).}
\thanks{This work was supported in part, by ONR Minerva under grant No. N$00014-18-1-2160$, and by NSF under grants No. SaTC-$1801611$ and  CPS-$1851588$.}}
\maketitle
 \thispagestyle{empty}
\pagestyle{empty}

\begin{abstract}
In this paper, we consider the problem of controlling  an underactuated system in unknown, and potentially adversarial environments. The emphasis will be on autonomous aerial vehicles, modelled by Dubins dynamics. The proposed control law is based on a variable integrator via online prediction for target tracking. To showcase the efficacy of our method, we analyze a pursuit evasion game between multiple autonomous agents. To obviate the need for perfect knowledge of the evader's future strategy, we use a deep neural network that is trained to approximate the behavior of the evader based on measurements gathered online during the pursuit.
\end{abstract}


\section{Introduction}
Output tracking in dynamical systems, such as robots, flight control, economics, biology, cyber-physical systems, is the practice of designing decision makers which ensure that a system's output  tracks a given signal \cite{devasia1996nonlinear,martin1996different}.


Well-known existing methods  for nonlinear output regulation and tracking include control techniques based on nonlinear inversions  \cite{Isidori90},  
  high-gain observers \cite{Khalil98}, and the framework of  
 model predictive control (MPC) \cite{allgower2012nonlinear,Rawlings17}. Recently a new approach has been proposed, based on 
  the Newton-Raphson flow for solving algebraic equations  \cite{Wardi17}. 
  Subsequently it has been tested on various  applications including controlling   an inverted pendulum,   and position control of platoons of mobile robotic vehicles \cite{Wardi18,shivam2018tracking}. While perhaps not as general as the aforementioned established techniques, it seems to hold out promise of efficient computations and large domains of stability.

 The successful deployment of complex control systems in real world applications increasingly  depends on their ability to operate on highly unstructured -- even adversarial -- settings, where \textit{a-priori} knowledge of the evolution of the environment is impossible to acquire. Moreover, due to the increasing interconnection between the physical and the cyber domains, control systems become more intertwined with human operators, making model-based solutions fragile to unpredictable. Towards that, methods that augment low-level control techniques with intelligent decision making mechanisms have been extensively investigated in \cite{saridis1983intelligent}.
Machine learning \cite{haykin2009neural,vrabie2013optimal}, offers a suitable framework to allow control systems to autonomously adapt by leveraging data gathered from their environment. To enable data-driven solutions for autonomy, learning algorithms use artificial neural networks (NNs); classes of functions that, due to properties that stem from their neurobiological analogy, offer adaptive data representations and prediction based on external observations.

NNs have been used extensively in control applications \cite{narendra1990identification}, both in open-loop and closed-loop fashion. In closed-loop applications, NNs have been utilized as dynamics approximators, or in the framework of reinforcement learning, in  enabling online solution of the Hamilton-Jacobi-Bellman equation \cite{vamvoudakis2010online}. However, the applicability of NNs in open-loop control objectives is broader, due to their ability to operate as classifiers, or as nonlinear function approximators \cite{bishop1995neural}.

 The authors of  \cite{narendra1990identification}  introduced NN structures for system identification as well as adaptive control. Extending the identification capabilities of learning algorithms, the authors of \cite{bhasin2013novel} introduce a robustification term that guarantees asymptotic estimation of the state and the state derivative. Furthermore, reinforcement learning has received increasing attention since the development of methods that solve optimal control problems for continuous time control systems online without the knowledge of the dynamics \cite{vamvoudakis2017q}.
Prediction has been in the forefront of research conducted on machine learning. Learning-based attack prediction was employed both in \cite{weber2009data} and \cite{alpcan2010network} in the context of cyber-security, and  \cite{pesch1995synthesis} utilized NNs to solve a pursuit evasion game by constructing both the evader's and the pursuer's strategies offline using pre-computed trajectories.  Recently, authors of this paper have applied NN for on-line model construction in a control application \cite{Kanellopoulos19}.

This paper applies an NN technique to the   pursuit-evasion problem
investigated  in  \cite{quintero2016robust}, which is more challenging than the problem addressed in \cite{Kanellopoulos19}.   The strategies of both pursuers and evader are based on  respective games. In  Ref. \cite{quintero2016robust},
the pursuers know the game of the evader ahead of time,  and an MPC technique is used to determine their trajectories. In this paper the pursuers  do not have an a-priori knowledge of the evader's game or its structure, and they employ an  NN
 in real time to identify its input-output mapping.  We use our tracking-control technique \cite{Wardi17} rather than MPC, and obtain similar results to \cite{quintero2016robust}.  Furthermore, the input to the  system has a lesser dimension that its output, 
and hence the control is underactuated. We demonstrate  a way of overcoming this limitation, which may have a broad  scope in applications.

The rest of the paper is structured as follows.
Section II describes our proposed control technique
and  some preliminary results on NN, and it  formulates the pursuers-evader problem. Section III describes results on model-based and learning-based strategies. Simulation results are presented in Section IV. Finally, Section V concludes the paper and discusses directions for future research.

\section{Preliminaries and Problem Formulation}

\subsection{Tracking Control Technique}\label{sec:tracking}
\begin{figure}
	\centering
	\includegraphics[width=2.8in]{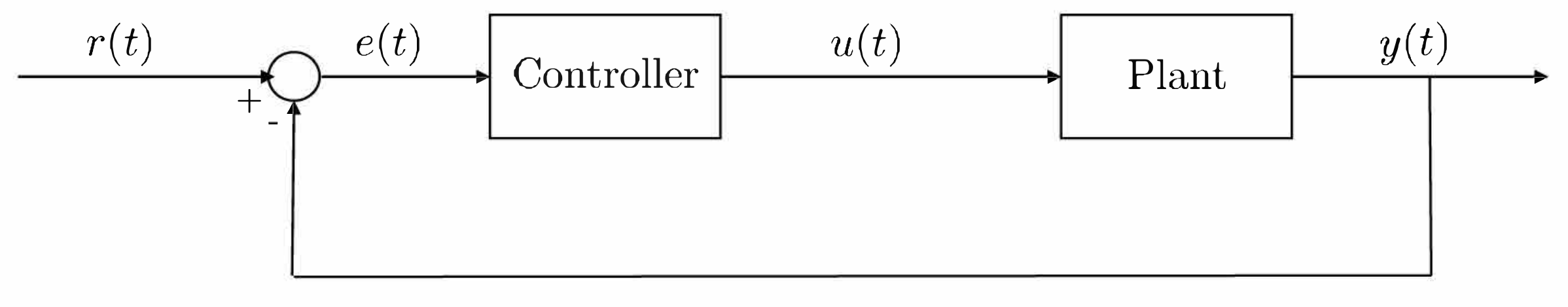}
	\caption{{ Basic control system scheme.}}\label{control_system}
\end{figure}

This subsection recounts results published in our previous work in which prediction-based output tracking was used for fully-actuated systems \cite{Wardi17, Wardi18,shivam2018tracking}. 
Consider a system as shown in Figure~\ref{control_system} with $r(t)\in \mathbb{R}^m$, $y(t)\in \mathbb{R}^m$,
$u(t)\in \mathbb{R}^m$, and $e(t):=r(t)-y(t)$. The objective of the controller is
to ensure that
\begin{equation}\label{track_error_eq}
\lim_{t\rightarrow\infty}||r(t)-y(t)||<\varepsilon,
\end{equation}
for a given (small) $\varepsilon\in\Real^+$. 

To illustrate the basic idea underscoring the controller, let us first assume that (i) The  plant subsystem is a memoryless nonlinearity of the form
\begin{equation}\label{eq:output}
    y(t)=g(u(t)),
\end{equation} 
for a continuously-differentiable function $g:\mathbb{R}^m\rightarrow \mathbb{R}^m$, 
and (ii) the target reference  $\{r(t):t\in[0,\infty)\}$ is a constant, $r(t)\equiv r$ for a given $r\in \mathbb{R}^m$.\footnote {Henceforth we will use the notation $\{x(t)\}$ For a generic signal $\{x(t),~t\in[0,\infty)\}$, to distinguish it from its value at a particular point $t$, $x(t)$.} These assumptions will be relaxed later.
In this case, the tracking controller is defined by the following
equation,
\begin{equation}\label{eq:controller}
    \dot{u}(t)=\Big(\frac{\partial g}{\partial u}(u(t))\Big)^{-1}\big(r-y(t)\big),
\end{equation}
assuming that the Jacobian matrix $\frac{\partial g}{\partial u}(u(t))$ is nonsingular at every point $u(t)$ computed by the controller via \eqref{eq:controller}. Observe that \eqref{eq:controller} defines the Newton-Raphson flow for solving the algebraic equation $r-g(u)=0$, and hence (see \cite{Wardi17,Wardi18})
the controller converges in the sense that 
$\lim_{t\rightarrow\infty}\big(r(t)-y(t)\big)=0$.
Next, suppose that the reference target is time-dependent, while keeping the assumption that the plant is a memoryless nonlinearity. Suppose that 
$\{r(t)\}$ is bounded, continuous, piecewise-continuously differentiable, and $\{\dot{r}(t)\}$ is bounded.
Define 
\begin{equation}\label{eq:rdot}
    \eta:=\lim\sup_{t\rightarrow\infty}||\dot{r}(t)||,
\end{equation}
then (see \cite{Wardi18}), with the  controller defined by  \eqref{eq:controller}, we have that
\begin{equation}\label{track_error_inf}
\lim_{t\rightarrow\infty}||r(t)-y(t)||\leq\eta.
\end{equation}

Note that Eqs.  \eqref{eq:output} and \eqref{eq:controller} together define the closed-loop system. Observe that the plant-equation \eqref{eq:output} is an algebraic equation while the controller equation \eqref{eq:controller} 
is a differential equation, hence the closed-loop system represents a dynamical system. Its stability, in the sense that $\{y(t)\}$ is bounded
whenever $\{r(t)\}$ and $\{\dot{r}(t)\}$ are bounded, is guaranteed by \eqref{track_error_inf} as long as the control
trajectory $\{u(t)\}$ does not pass through a point $u(t)$ where the Jacobian matrix $\frac{\partial g}{\partial u}(u(t))$ is singular.

Finally, let us dispense with the assumption that the plant subsystem is a memoryless nonlinearity.
Instead, suppose that it is   a dynamical system modeled by the following two equations,
\begin{align}
    \dot{x}(t)&=f(x(t),u(t)),~~x(0):=x_{0}\label{eq:state_eqn}\\
    y(t)&=h(x(t)), \label{eq:general_output}
\end{align}
where the state variable $x(t)$ is in $\mathbb{R}^n$, and the functions $f:\mathbb{R}^n\times \mathbb{R}^m\rightarrow \mathbb{R}^n$ and $h:\mathbb{R}^n\rightarrow \mathbb{R}^m$ satisfy the following
assumption.
\begin{assumption}
(i). The function $f:\mathbb{R}^n\times \mathbb{R}^m\rightarrow \mathbb{R}^n$ is continuously differentiable, and for every compact set $\Gamma\subset \mathbb{R}^m$
there exists $K\in\Real^+$ such that, for every $x\in \mathbb{R}^n$ and $u\in\Gamma$,
$||f(x,u)||\leq K\big(||x||+1\big)$.
(ii). The function $h:\mathbb{R}^n\rightarrow \mathbb{R}^m$ is continuously differentiable. \frqed
\end{assumption}
This assumption ensures that whenever the control signal $\{u(t)\}$  is bounded and continuous, the state equation \eqref{eq:state_eqn} has a unique solution $x(t)$ on the interval $t\in[0,\infty)$.

In this setting, $y(t)$ is no longer a function of $u(t)$, but rather of 
$x(t)$ which is a function of $\{u(\tau):\tau<t\}$. Therefore \eqref{eq:output} is no longer valid, and hence the controller cannot be defined by \eqref{eq:controller}. To get around this conundrum we pull the feedback not from the output $y(t)$ but from a predicted value thereof. Specifically, fix the look-ahead time $T\in\Real^+$, and suppose that at time $t$ the system computes a prediction of $y(t+T)$, denoted by $\tilde{y}(t+T)$. Suppose also that $\tilde{y}(t+T)$ is a function of $(x(t),u(t))$,  hence can be written as
$\tilde{y}(t+T)=g(x(t),u(t))$,
where the function $g:\mathbb{R}^n\times \mathbb{R}^m\rightarrow \mathbb{R}^m$ is continuously differentiable. 

Now the feedback law is defined by the following equation,
\begin{equation}\label{eq:control_full}
    \dot{u}(t)=\Big(\frac{\partial  g}{\partial u}(x(t),u(t))\Big)^{-1}\big(r(t+T)-g(x(t),u(t))\big). 
\end{equation}
The state equation \eqref{eq:state_eqn} and control equation \eqref{eq:control_full} together define the closed-loop system. This system can be viewed as an $(n+m)$-dimensional dynamical system with the state variable $(x(t)^{\textrm{T}},u(t)^{\textrm{T}})^{\textrm{T}}\in \mathbb{R}^{n+m}$ and input $r(t)\in \mathbb{R}^m$. We are concerned with a variant of Bounded-Input-Bounded-State (BIBS) stability whereby if $\{r(t)\}$ and $\{\dot{r}(t)\}$ are
bounded, $\{x(t)\}$ is bounded as well. Such stability 
no-longer can be taken for granted as in the case where the plant is a memoryless nonlinearity. 

We remark that a larger $T$ means larger prediction errors, and these  translate into larger asymptotic tracking errors. On the other hand, an analysis of various second-order systems in \cite{Wardi17} reveals that they all were unstable if $T$ is too small, and stable if $T$ is large enough. 
It can be seen that, a requirement for a restricted prediction error can stand in contradiction with the stability requirement. This issue was resolved  by speeding up the controller in the following manner.
Consider  $\alpha>1$, and modify \eqref{eq:control_full} by multiplying its right hand side by $\alpha$, resulting in the following control equation: 
\begin{equation*}
    \dot{u}(t)=\alpha\Big(\frac{\partial  g}{\partial u}(x(t),u(t))\Big)^{-1}\big(r(t+T)-g(x(t),u(t))\big). 
\end{equation*}
It was verified in \cite{Wardi17,Wardi18,shivam2018tracking}, that regardless of the value of $T\in\Real^+$, a large-enough $\alpha$ stabilizes the closed-loop system.\footnote{This statement seems to have a broad scope, and does not require the plant to be a minimum-phase system.}  Furthermore, if the closed-loop system is stable 
 then the following bound holds,
\begin{equation}\label{eq:error_alpha}
    \lim\sup_{t\rightarrow\infty}||r(t)-\tilde{y}(t)||\leq\frac{\eta}{\alpha},
\end{equation}
where $\eta$ is defined by \eqref{eq:rdot}. Thus, a large gain $\alpha$ can stabilize the closed-loop system and reduce the asymptotic tracking error.

\subsection{Problem Formulation}
In an attempt to broaden the application scope of the control algorithm, underactuated systems such as the fixed-wing aircraft are explored, which are widely used in the domain of aerospace engineering. 
The behavior of a fixed wing aircraft at constant elevation can be approximated by a planar Dubins vehicle with $3$ states \cite{lavalle2006planning} $\forall t\geq0$,
\begin{align*}
\dot{z}_1^p(t)&=V^{p}\cos\theta^p(t)\text{, }\\
\dot{z}_2^p(t)&=V^{p}\sin\theta^p(t) \text{, }\\
\dot{\theta}^p(t)&=u(t),
\end{align*}
where $( z^p_1(t),z^p_2(t))^{\textrm{T}}$ denotes the planar position of the vehicle, $\theta^p(t)$ its heading and $u(t)$ the angular acceleration, constrained as, $\norm{u} \leq u_{\textrm{max}}$. The input saturation enforces a minimum turning radius equal to $V_0/u_{\textrm{max}}$.   For testing the efficacy of the controller for  the underactuated system, henceforth referred to as the pursuer,  it is tasked with tracking an evading vehicle, modeled as a single integrator,  with dynamics as follows:
\begin{equation*}\frac{\textrm{d}}{\textrm{d}t}
    \begin{bmatrix}
    z_1^\textrm{e}(t)\\[0.21em]
    z_2^\textrm{e}(t)
    \end{bmatrix} =
    \begin{bmatrix}
    V^{\textrm{e}}\cos\theta^{\textrm{e}}\\[0.2em]
    V^{\textrm{e}}\sin\theta^{\textrm{e}}\\[0.2em]
    \end{bmatrix},
\end{equation*}
where $(z_1^e(t),z_2^e(t))^{\top}$ denote the planar position of the evader, and $V^e$ is its speed. 
We consider two cases; one where the evader is agnostic to the pursuer and follows a known trajectory and the other where the the evader is adversarial in nature and its trajectory is not known to the pursuer. 
The next section will provide two  solutions for the problem of estimating the evader's trajectory based, respectively, on a model-based approach and a learning-based approach.

\section{Predictive Framework}
\subsection{Model-Based Pursuit Evasion}
The considered system is underactuated because the  pursuer's position, $(z_{1}^p(t),z_{2}^p(t))^{\top}$,  is two-dimensional while  it is controlled by an one-dimensional variable, $u(t)$. This raises a problem since the application of the proposed tracking technique requires the control variable and system's output to have the same dimension. To get around this difficulty,  we define
a suitable  function $F:R^2\rightarrow R^+$ and   set $g(x(t),u(t)):=\int_t^{t+T}F(\tilde{y}^p(\tau)-\tilde{y}^e(\tau))\textrm{d}\tau$ where
 $\tilde{y}^p(\tau)$ and $\tilde{y}^e(\tau)$ are the predicted position of the pursuer and the evader at time $\tau$; we  apply  the Newton-Raphson flow to the equation $g(x(t),u(t))=0$.  The modified controller becomes
\begin{equation}\label{u_dot}
\dot{u}(t)=-\alpha\Big(\frac{\partial  g}{\partial u}(x(t),u(t))\Big)^{-1}\big(g(x(t),u(t))\big) ,\ t\geq0.
\end{equation}
 Since $g(x,u)$ is a scalar,  the modified algorithm works similar to the base case. 

Assume general nonlinear system dynamics as in \eqref{eq:state_eqn} with output described in \eqref{eq:general_output}.  The predicted state trajectory is computed by holding the input to a constant value over the prediction horizon, given by the following differential equation:

\begin{equation}\label{predicted_state}
\dot{\xi}(\tau)=f(\xi(\tau),u(t)), ~\tau\in [t,t+T],    
\end{equation}
with the initial condition $\xi(t)=x(t) $ as shown in \cite{Wardi17}. The predicted output at $\tau$ is $\tilde{y}^p(\tau)=h(\xi(\tau))$. Furthermore, by taking the partial derivative of  \eqref{predicted_state} with respect to u(t), we obtain
\begin{equation}\label{predicted_derivative}
\dot{\frac{\partial \xi}{\partial u}}(\tau)=\frac{\partial f}{\partial \xi}(\xi(\tau),u(t))\frac{\partial \xi}{\partial u}(\tau)+\frac{\partial f}{\partial u}(\xi(\tau),u(t)),
\end{equation}
with the initial condition ${\frac{\partial \xi}{\partial u}}(t)=0$. The above is a differential equation in ${\frac{\partial \xi}{\partial u}}(\tau); ~\tau \in [t,t+T]$ and  \eqref{predicted_state} and \eqref{predicted_derivative} can be solved numerically. 
 Finally, the values of $g(x,u)$ and $\frac{\partial g}{\partial u}(x,u)$ can be substituted in  \eqref{u_dot} to  get the control law.


In the next section, results are presented for an agnostic as well as an adversarial pursuer- evader system. However, as mentioned above, in the adversarial problem formulation, the trajectory of the evader is not known in advance, which can be overcome in two ways. 

In the first approach, the pursuer(s) use game theory to predict the approximate direction of evasion. As mentioned in \cite{isaacs1999differential}, in the case of single pursuer, the evader's optimal strategy is to move along the line joining the evader and pursuer's position, if the pursuer is far enough. When the distance between the pursuer and the evader reduces to the turning radius of the pursuer, the evader switches strategies and enters into the non-holonomic constraint region of the pursuer. This can be represented as follows:
\begin{equation}\label{evasion}
    \theta_E= \begin{cases} 
                \arctan\bigg({\frac{\vphantom{z_{2_p}}z_2^e(t)-z_2^p(t)}{z_1^e(t)-\vphantom{z_1^{p^p}(t)}z_1^{p}(t)}}\bigg), & d > R_P, \\ \\
                \arctan\bigg({\frac{\vphantom{z_{2_p}}z_2^e(t)-z_2^p(t)}{z_1^e(t)-\vphantom{z_1^{p^p}(t)}z_1^{p}(t)}}\bigg) \pm \pi/2, & d\leq R_P.
   \end{cases}
\end{equation}
Here $\theta_E$ is the expected evasion angle of the evader and $d$ is the distance between the pursuer and evader, 

If there are multiple pursuers, it is assumed that the evader follows the same strategy by considering only the closest pursuer. It is notable that this will not provide the pursuers a correct prediction of the evader's motion as they do not know about the goal seeking behavior mentioned above. However, it gives a good enough approximation of the pursuer's motion that the algorithm can be used for tracking. 

The second approach involves learning the evader's behavior over time using NN. The pursuers take their positions and the position of the evader as input and the NN gives the estimated evasion direction as the output after training. 

To showcase the efficacy of our method, we consider a pursuit evasion problem, involving multiple pursuing agents. Such problems are typically formulated as zero-sum differential games \cite{isaacs1999differential}. Due to the difficulty of solving the underlying Hamilton-Jacobi-Isaacs (HJI) equations \cite{basar1999dynamic} of this problem, we shall utilize the method described in \ref{sec:tracking} to approximate the desired behavior.
Furthermore, we show that augmenting the controller with learning structures in order to tackle the pursuit evasion problem without explicit knowledge of the evader's behavior is straightforward.

In order to formulate the pursuit evasion problem, we define a global state space system consisting of the dynamics of the pursuers and the evader. For ease of exposition, the analysis will focus on the $2$-pursuer, $1$-evader problem, since extending the results to multiple pursuers is straightforward.

The global state dynamics become,
\begin{equation}\label{eq:nonlin_dynam}
\frac{\textrm{d}}{\textrm{d}t}
    \begin{bmatrix}
    z_1^{\textrm{p}_1}(t)\\[0.21em]
    z_2^{\textrm{p}_1}(t)\\[0.21em]
    \theta^{\textrm{p}_1}(t)\\[0.21em]
    z_1^{\textrm{p}_2}(t)\\[0.21em]
    z_2^{\textrm{p}_2}(t)\\[0.21em]
    \theta^{\textrm{p}_2}(t)\\[0.21em]
    z_1^\textrm{e}(t)\\[0.21em]
    z_2^\textrm{e}(t)
    \end{bmatrix} =
    \begin{bmatrix}
    V^{\textrm{p}_1}\cos\theta^{\textrm{p}_1}\\[0.2em]
    V^{\textrm{p}_1}\sin\theta^{\textrm{p}_1}\\[0.2em]
    u^\textrm{p}_1\\[0.2em]
    V^{\textrm{p}_1}\cos\theta^{\textrm{p}_2}\\[0.2em]
    V^{\textrm{p}_2}\sin\theta^{\textrm{p}_2}\\[0.2em]
    u^\textrm{p}_2\\[0.2em]
    V^{\textrm{e}}\cos\theta^{\textrm{e}}\\[0.2em]
    V^{\textrm{e}}\sin\theta^{\textrm{e}}\\[0.2em]
    \end{bmatrix},
\end{equation}
where the subscripts indicate the autonomous agent. For compactness, we denote the global state vector as $x(t)\in \mathbb{R}^8$, the pursuers' control vector $u(t) \in \mathbb{R}^2$, and the nonlinear mapping described by the right-hand side of \eqref{eq:nonlin_dynam}. Thus, given the initial states of the agents $x_0\in\mathbb{R}^8$, the evolution of the pursuit evasion game is described by 
$\dot{x}(t) = f(x(t),u,u_\textrm{e})\text{, }x(0)=x_0\text{, }t\geq 0$.

Subsequently, this zero-sum game can be described as a minimax optimization problem through the cost index,
\begin{align}\label{eq:cost}
J(x,u,u_\textrm{e}) &= \int_{0}^{\infty} e^{-\gamma t}L(x)\textrm{d}t \nonumber\\&:= \int_0^\infty e^{-\gamma t}\bigg(\beta_1(d^2_1+d_2^2)+\beta_2\frac{d^2_1d^2_2}{d^2_1+d^2_2}\bigg)\textrm{d}t,
\end{align}
where $d_i=\sqrt{(z_1^i-z^\textrm{e})^2+(z_2^i-z_2^\textrm{e})^2}$, $i\in\lbrace \textrm{p}_1,\textrm{p}_2\rbrace$ is the distance between the $i$-th pursuer and the evader, $\beta_1,\ \beta_2\in\Real^+$ are user defined contants, and $\gamma\in\Real^+$ is a discount factor. The first term ensures that the pursuers remain close to the evader, while the second term encourages cooperation between the agents. The cost decreases exponentially to ensure that the integral has a finite value in the absence of equilibrium points. 

Let $V(x):\mathbb{R}^8\rightarrow\mathbb{R}$ be a smooth  function quantifying the value of the game when specific policies are followed starting from state $x(t)$.
Then, we can define the corresponding Hamiltonian of the game as,
\begin{equation}\label{eq:ham}
H\big(x,u,u_e,\frac{\partial V}{\partial x}\big) = L(x) + \frac{\partial V}{\partial x}^{\textrm{T}}f(x,u,u_e) + \gamma V.
\end{equation}

The optimal feedback policies $u^\star(x)$, $u^\star_e(x)$ of this game are known to constitute a saddle point \cite{basar1999dynamic} such that,
\begin{align}
u^\star(x) = \arg\min_u H(x,u,u_e), \label{eq:opt_purs}\\
u_e^\star(x) = \arg\max_{u_e} H(x,u,u_e). \label{eq:opt_evade}
\end{align}
Under the optimal policies \eqref{eq:opt_purs},\eqref{eq:opt_evade}, the HJI equation is satisfied,
\begin{equation}\label{eq:HJI}
H\big(x,u^\star,u_e^\star,\frac{\partial V}{\partial x}^\star\big) = 0.
\end{equation}
Evaluating the optimal pursuit policies, yields the singular optimal solutions described by,
$V_{\theta_{p1}}u_1 = V_{\theta_{p2}}u_2 =0$,
where $V_{x_i}$ is the partial derivative of the value function with respect to the state $x_i$, calculated by solving \eqref{eq:HJI}.
To obviate the need for bang-bang control, as is derived by \eqref{eq:opt_purs} and \eqref{eq:opt_evade} we shall employ the predictive tracking technique described in Section \ref{sec:tracking} to derive approximate, easy to implement, feedback controllers for the pursuing autonomous agents. Furthermore, by augmenting the predictive controller with learning mechanisms, the approximate controllers will have no need for explicit knowledge of $u_e^\star(x)$, the evader's policy.

The following theorem presents bounds on the optimality loss induced by the use of the look-ahead controller approximation.

\begin{theorem}
Let the pursuit evasion game evolve according to the dynamics given by \eqref{eq:nonlin_dynam}, where the evader is optimal with respect to \eqref{eq:cost} and the pursuers utilize the learning-based predictive tracking strategy given \eqref{u_dot}. Then, the tracking error of the pursuers and the optimality loss due to the use of the predictive controller are bounded if $\exists \bar{\Delta}\in\Real^+$, such that, $\Delta(x(t),\hat{u}(t),\hat{u}(t)_\textrm{e}) \leq \bar{\Delta},~\forall t\geq0$,
where $
    \Delta(x,\hat{u},\hat{u}_\textrm{e}) = V_{x_\textrm{e}}v_\textrm{e}(\cos\hat{u}_\textrm{e}-\cos u^\star_\textrm{e})+V_{y_\textrm{e}}v_\textrm{e}(\sin \hat{u}_\textrm{e}-\sin u^\star_\textrm{e}) + V_{\theta_\textrm{p}1}(u_1^\star-\hat{u}_1)+V_{\theta_\textrm{p}2}(u_2^\star-\hat{u}_2),
$
with $V_{\xi}$ denoting the partial derivative of the game value with respect to the state component $\xi(t)$.
\end{theorem}

Proof: Consider the Hamiltonian function when the approximate controller, denoted $\hat{u}(t)$ and the NN-based prediction of the evader's policy, $\hat{u}_\textrm{e}(t)$ are used,
\begin{align}\label{eq:subopt_hamiltonian}
    H(x,\hat{u},\hat{u}_\textrm{e}) = L(x) + \big(\frac{\partial V}{\partial x}\big)^\textrm{T} f(x,\hat{u},\hat{u}_\textrm{e}) +\gamma V.
\end{align}
Taking into account the nonlinear dynamics of the system \eqref{eq:nonlin_dynam}, one can rewrite \eqref{eq:subopt_hamiltonian} in terms of the optimal Hamiltonian as,$H(x,\hat{u},\hat{u}_\textrm{e}) = H(x,u^\star,u^\star_\textrm{e}) + \Delta(\hat{u},\hat{u}_\textrm{e})$,
where $H(x,u^\star,u^\star_\textrm{e})=0$ is the HJI equation that is obtained after substituting \eqref{eq:opt_purs} and \eqref{eq:opt_evade} in \eqref{eq:ham}.
Now, take the orbital derivative of the value function along the trajectories using the approximate controllers as,
$
    \dot{V} = \big(\frac{\partial V}{\partial x}\big)^\textrm{T} f(x,\hat{u},\hat{u}_\textrm{e}).
$
Substituting \eqref{eq:subopt_hamiltonian} yields
$
    \dot{V} = -L(x) - \gamma V + \Delta(x,\hat{u},\hat{u}_\textrm{e}).
$
Thus, since $L(x)> 0$, $\forall x \in \mathbb{R}^8\setminus \lbrace 0 \rbrace$,
\begin{align*}
    \dot{V} < -\gamma V + \Delta(x,\hat{u},\hat{u}_\textrm{e}) \Rightarrow\dot{V} < -\gamma V + \bar{\Delta}.
\end{align*}

Hence for $V\geq\bar{\Delta}/\gamma$, we have $\dot{V}\leq0$. Thus $\lbrace x\in \mathbb{R}^8~|~ V(x)\leq \bar{\Delta}/\gamma \rbrace$ is a forward invariant set, which implies that the tracking error and the optimality loss over any finite horizon is bounded.
\frQED

\begin{remark}
Note that we do not use optimal control or MPC to solve the pursuit evasion problem. Instead, the controller is governed by \eqref{u_dot}, which is simple to implement and has low computational complexity.
\frqed
\end{remark}

\subsection{Deep Learning-Based Pursuit Evasion}

A deep NN, consisting of $L > 2$ hidden layers, describes a nonlinear mapping between its input space $\mathbb{R}^n$ and output space $\mathbb{R}^p$. 
Each layer receives the output of the previous layer as an input and, subsequently, feeds its own output to the next layer. Each layer's output consists of the weighted sum of its input alongside a bias term, filtered through an application-specific activation function \cite{haykin2009neural}.

Specifically, let $\mathbb{R}^{n_l}$ be the input space of a specific layer, and $\mathbb{R}^{p_l}$ the corresponding output space. Then the layer's output is,
\begin{equation*}
Y_i(x) = \sigma\bigg(\sum_{j=1}^{n_l} v_{ij}X_j + v_{i0}\bigg)\text{, } i = 1,2,\dots,p_l, 
\end{equation*}
where $X^\prime   = \begin{bmatrix}X_1 & \dots & X_{n_l} \end{bmatrix}^\textrm{T} \in \mathbb{R}^{n_l}$ is the input vector, gathered from training data or from the output of previous layers, $v_{ij}\in \mathbb{R}$ is a collection of $n_l$ weights for each layer, $v_{i0} \in \mathbb{R}$ the bias term and $\sigma: \mathbb{R}^{n_l} \rightarrow \mathbb{R}$ is the layer's activation function. 
We note that it is typical to write the output of layer compactly, with slight abuse of notation, as,
\begin{equation}\label{eq:NN}
Y = \sigma(W^\textrm{T} \sigma^\prime(X)),
\end{equation}
where $Y = \begin{bmatrix} Y_1 &\dots & Y_{p_l} \end{bmatrix} \in \mathbb{R}^{p_l}$, $W = \begin{bmatrix} v_{ij} \end{bmatrix}\in \mathbb{R}^{(n_l+1)\times p_l}$ and $\sigma^\prime:\mathbb{R}^{n_l^\prime} \rightarrow \mathbb{R}^{n_l}$ is the activation function of the previous layer, taking as input the vector $X=\begin{bmatrix}{X^\prime}^{\textrm{T}}& 1 \end{bmatrix}^{\textrm{T}}$ .

It is known \cite{lewis1998neural}, that two-layer NNs possess the universal approximation property, according to which, any smooth function can be approximated arbitrarily close by an NN of two or more layers.
Let $\mathbb{S}\subset \mathbb{R}^n$ be a simply connected compact set and consider the nonlinear function $\kappa : \mathbb{S} \rightarrow \mathbb{R}^p$. Given any $\epsilon_b \geq 0$, there exists a NN such structure such that,
\begin{equation*}
\kappa(x) = \sigma\big(W^\textrm{T} \sigma^\prime(x)\big) + \epsilon\text{, } \forall x \in \mathbb{S},
\end{equation*}
where $\|\epsilon\| \leq \epsilon_b$. We note that, typically, the activation function of the output layer $\sigma(\cdot)$ is taken to be linear.

Evaluating the weight matrix $W$ in a network is the main concern of the area of machine learning. In this work, we employ the gradient descent based backpropagation algorithm.
Given a collection of $N_d$ training data, stored in the tuple $\lbrace x_k,\kappa_k \rbrace_{k}$, where $x_k \in \mathbb{R}^n$, $\kappa_k \in \mathbb{R}^p$, $\forall k =1,\dots,N_d$, we denote the output errors as
$
r_k = \kappa(x_k) - \kappa_k.
$
Then, the update equation for the weights at each optimization iteration $t_k$ is given by,
\begin{align}\label{eq:NN_tune}
  w_{ij}(t_k+1) = w_{ij}(t_k) - \eta \frac{\partial (r_k^\textrm{T}r_k)}{\partial w_{ij}}, ~\forall t_k \in \mathbb{N},  
\end{align}
where $\eta\in\Real^+$ denotes the learning rate. We note that the update index $t_k$ need not correspond to the sample index $k$, since different update schedules leverage the gathered data in different ways \cite{lewis1998neural}.
It can be seen that in order for the proposed method to compute the pursuers' control inputs, an accurate prediction of the future state of the evader is required. However, this presupposes that the pursuers themselves have access to the evader's future decisions; an assumption that is, in most cases, invalid.
Thus, we augment the pursuers' controllers with a NN structure, that learns to predict the actions of the evader, based on past recorded data.

Initially, we assume that the evader's strategy is computed by a feedback algorithm, given her relative position to the pursuers. This way, the unknown function we wish to approximate is $f:\mathbb{R}^{2N} \rightarrow \mathbb{R}^2$, with,
$
u^e = f(\delta z_1^{{p}_1}, \delta z_2^{p_1} ,\dots, \delta z_1^{p_N},\delta z_2^{p_N}), $
where, $(\delta z_1^{p_i},\delta z_2^{p_i})$ denote the distance of pursuer $i$ to the evader in the X and Y axes, respectively.
In order to train the network, we let the pursuers gather data regarding the fleet's position with respect to the evader, as well as her behavior over a predefined time window $T_l > 0$.

\begin{remark}
Increasing the time window $T_l$ will allow the pursuers to gather more training data for the predictive network. However, this will not only increase the computational complexity of the learning procedure, but will make the pursuers more inert to sudden changes in the evader's behavior. Simulation results corroborate our choice of training parameters. \frqed 
\end{remark}

Subsequently, we denote by $\hat{u}^e(x)$, the current prediction function for the evader's strategy, i.e., $\hat{u}^e(x) = \sigma\big(\hat{W}^\textrm{T}\hat{\sigma}^\prime(\chi)\big)$,
where $\chi = \begin{bmatrix} \delta z_1& \delta y_1 &\dots& \delta x_N & \delta y_N \end{bmatrix} \in \mathbb{R}^{2N}$, $\hat{W}$ denotes the current weight estimate of the NNs output layer, and $\hat{\sigma}^\prime(\cdot)$ is the current estimate of the hidden layers, parametrized by appropriate hidden weights.

\begin{remark}
While the learning algorithm for the evader's behavior operates throughout the duration of the pursuit, thus making the approximation weights time-varying, we suppress their explicit dependence on time since the process is open-loop, in the sense that the system is learning in batches, rather that in a continuous fashion. \frqed
\end{remark} 
\begin{algorithm}[b]
    \caption{Deep Learning-Based and Predictive Pursuit Evasion}
    \textbf{Inputs:} ${X_{P_i}(t)}$, $\forall i\lbrace 1,\dots,N\rbrace$, $X_E(t)$ and evasion strategy approximation weights $W$.\\ 
    \textbf{Output:} ${u_{P_i}(t)}$, $\forall i\lbrace 1,\dots,N\rbrace$.
    \begin{algorithmic}[1]
        \State Compute $(\delta x_i,\delta y_i)$, $i\in\lbrace1,\dots,N\rbrace$.
            \State Predict evader's future behavior via \eqref{eq:NN}.
        \State Train NN as in \eqref{eq:NN_tune}.
        \State Predict evader's future state as $\tilde{X}_E(t+T)=X_E(t)+[V_E\cos{\theta_E} ~~V_E\sin{\theta_E}]^{\textrm{T}}T$.
        \State Propagate pursuer dynamics to get $\tilde{X}_P(t+T)$.
        \State Computed current Newton flow parameters using \eqref{g_multi_pursuer}.
        \State Computed control dynamics $\dot{u}_{P_i}(t)$ from \eqref{eq:controller}.
        \State Propagate actual system evolution using \eqref{eq:nonlin_dynam}.
        \State Append current distances $(\delta x_i,\delta y_i)$ to a stack of previous observations.
        \State Update evader prediction network through \eqref{eq:NN_tune}.
    \end{algorithmic}
\end{algorithm}    
\section{Simulation Results}
This section presents results for the problems briefly described in the previous section. First, the agnostic evader case is considered followed by the adversarial case. For the second case, single and multiple pursuer systems are considered separately. The controller is implemented on a Dubins vehicle. For the purpose of tracking, we define the system output to be $y^i=\begin{bmatrix}z_1^i&z_2^i\end{bmatrix}^\textrm{T}$, $i \in \lbrace p_1,p_2,e\rbrace$.
\vspace{-1mm}
\subsection{Single Pursuer - Agnostic Target}
In this subsection, the controller is tested on a Dubins vehicle with the task of pursuing an agnostic target moving along a known trajectory. Since the vehicle has a constant speed and an input saturation is enforced, it has an inherent minimum turning radius. For this simulation, we set $V^p=2$~m/s and the input saturation is first set to $\frac{\pi}{2}$~rad/s and then to $2{\pi}$~rad/s. The evader moves along two semicircular curves with a constant speed which is less than $V^p$.

\begin{figure}[!ht]
\vspace{6pt}
\begin{center}
	\includegraphics[width=0.9\linewidth]{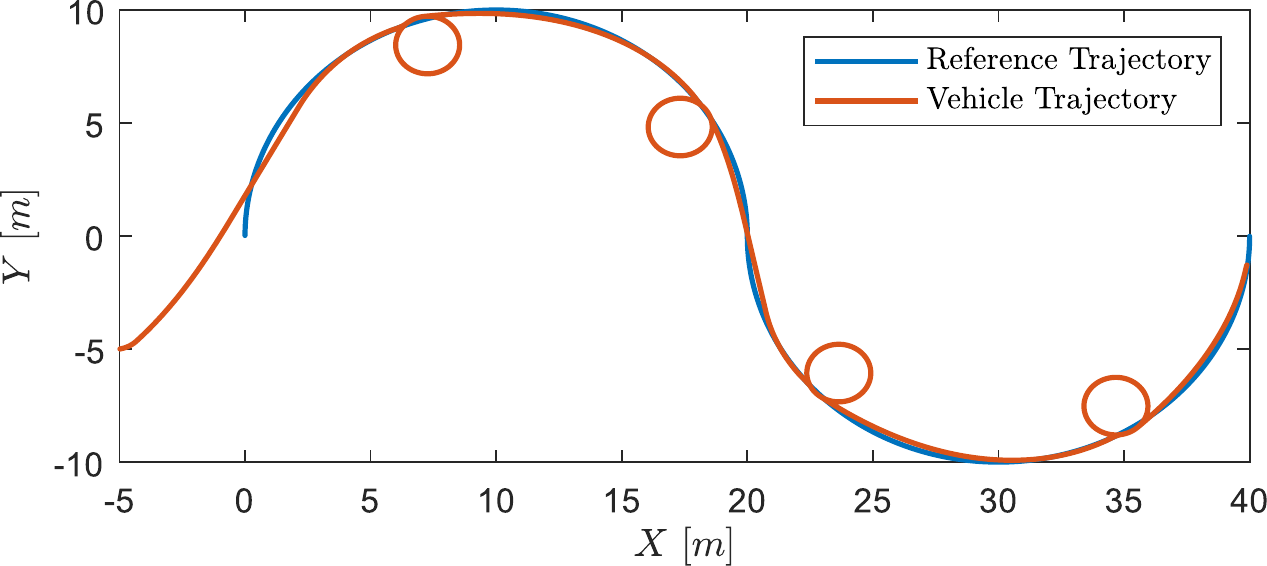}
	\vspace{-8pt}	
	\caption{ Agnostic evader with a large turning radius.}\label{trajectory_R}
\end{center}
\begin{center}
	\includegraphics[width=0.9\linewidth]{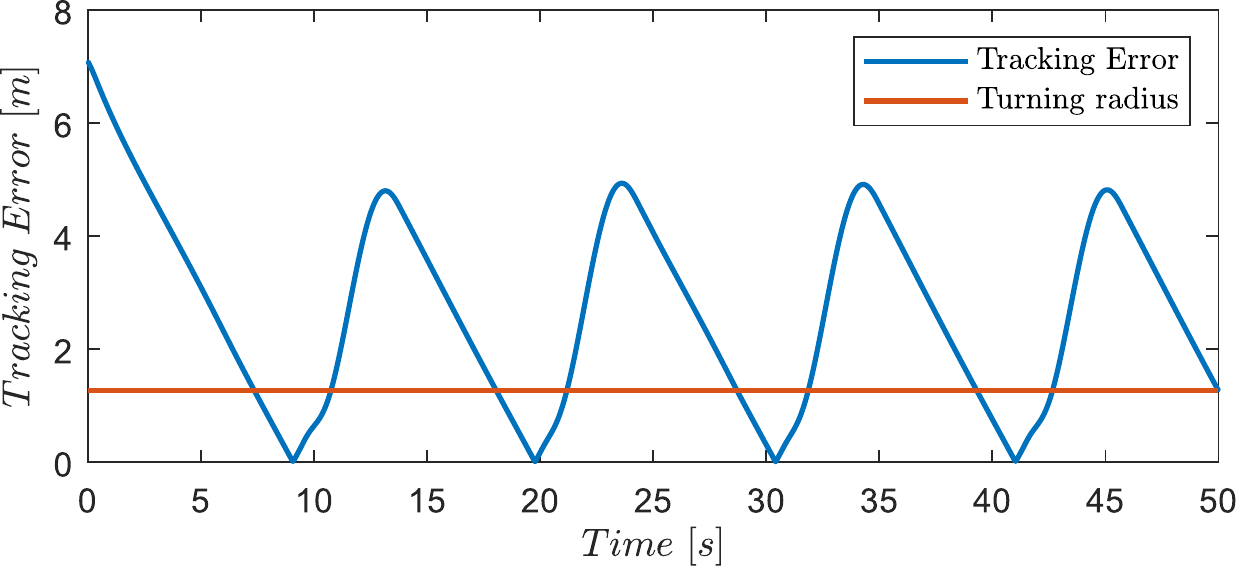}
	\vspace{-8pt}
	\captionof{figure}{{Evolution of an  agnostic evader tracking error with a large turning radius.}}\label{error_R}
\end{center}
\begin{center}
	\includegraphics[width=0.9\linewidth]{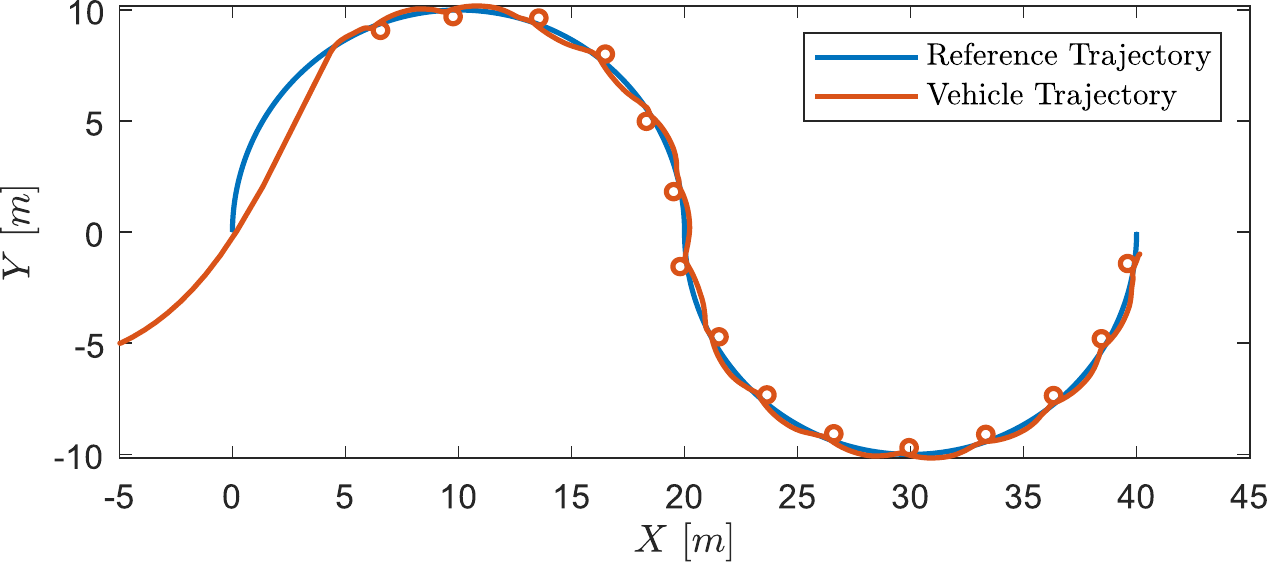}
	\vspace{-8pt}
	\captionof{figure}{{Agnostic evader  with a small turning radius.}}\label{trajectory_r}
\end{center}
\begin{center}
	\includegraphics[width=0.9\linewidth]{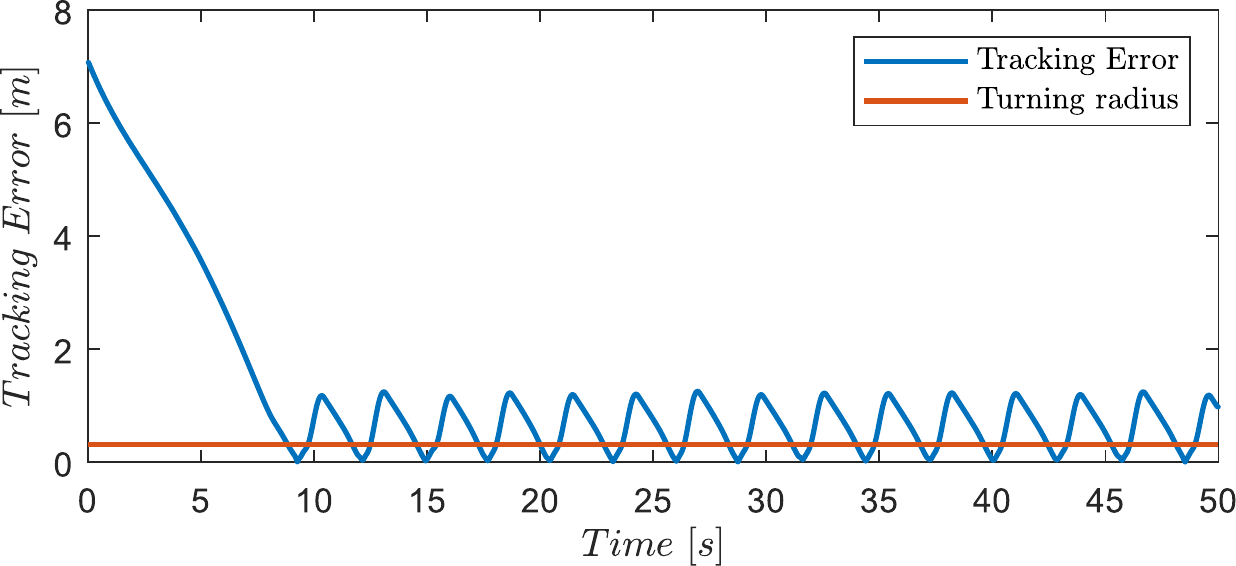}
	\vspace{-8pt}
	\captionof{figure}{{ Evolution of the agnostic evader tracking error with a small turning radius.}}\label{error_r}
\end{center}
\end{figure}

As a consequence, when the pursuer catches up to the evader, it overshoots and has to go around a full circle to again start tracking. Naturally, lower turning radius translates to better tracking as the vehicle can make ``tighter'' turns. This can be seen when comparing the trajectories of the vehicle in Figure~\ref{trajectory_R} with Figure~\ref{trajectory_r}. For the same trajectory of the evader, the tracking performance is far better in the second case. Once the pursuer catches up to the target, the maximum tracking error in the first case is approximately $4$ meters and only $1$ meter in the second case, shown in Figures~\ref{error_R} and \ref{error_r}. This is consistent with the fact that the ratio of the turning radii is $4:1$.

\subsection{Single Pursuer - Adversarial Evader}
The pursuer is again modelled as a Dubins vehicle, while the evader is modelled as a single integrator with a maximum velocity less than the speed of the pursuer. Hence, while the pursuer is faster, the evader is more agile, and can instantly change its direction of motion. In this and subsequent cases, the evader is considered adversarial in nature and uses game theory to choose evasion direction.

Let $y^p(t)$ and $y^e(t)$ be the position vector of the pursuer and evader respectively at time $t$. First, the pursuer makes an estimate of the optimal evasion direction based on the relative position of the evader and itself at time $t$ using \eqref{evasion}. Assuming this direction of evasion to be fixed over the prediction window from $t$ to $t+T$ gives the predicted position of the evader at all time instances in this interval, denoted as $\tilde{y}^e(\tau), \tau \in [t, t+T]$. Next, the pursuer estimates its own predicted position if its input is kept constant, called $\tilde{y}^p(\tau),\tau \in [t, t+T]$. Finally, $g(t)$ is set as $||\tilde{y}^e(t+T)-\tilde{y}^p(t+T)||^2$ and the value of $\frac{\partial g}{\partial u}(x(t),u(t))$ ($x(t)$ being the ensemble vector of the states of the pursuer and the evader) is used to compute  the input differential equation \eqref{u_dot}.

\begin{figure}[!ht]
\vspace{6pt}
\begin{center}
	\includegraphics[width=0.9\linewidth]{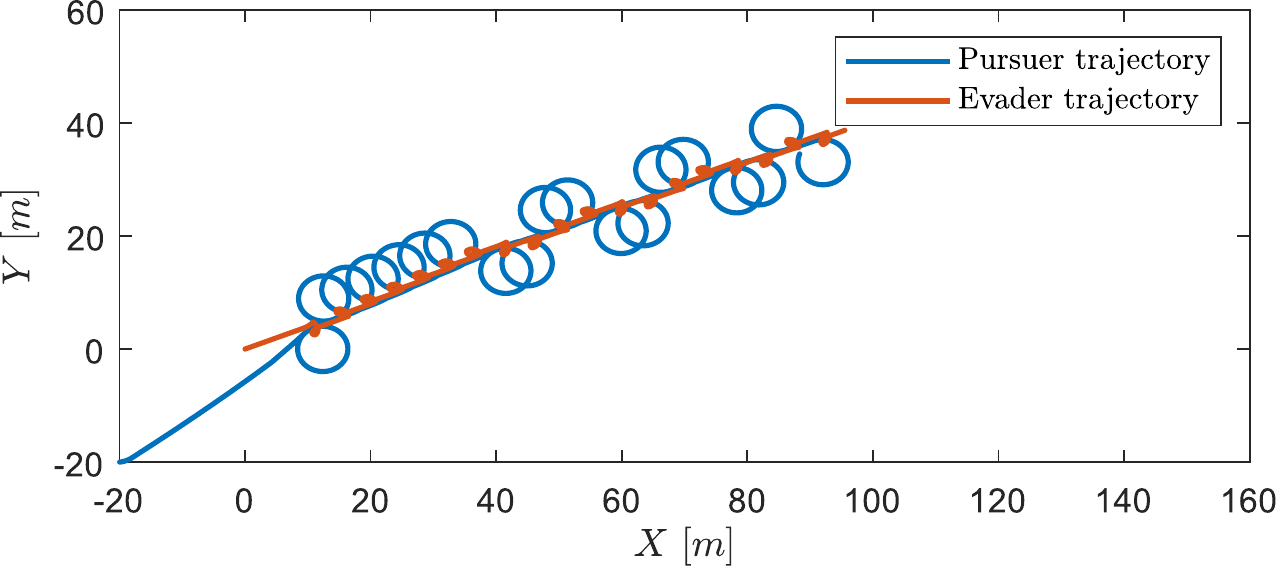}
	\vspace{-8pt}	
	\caption{ Trajectories for a single pursuer-evader system.}\label{trajectory_pursuer}
\end{center}
\begin{center}
	\includegraphics[width=0.9\linewidth]{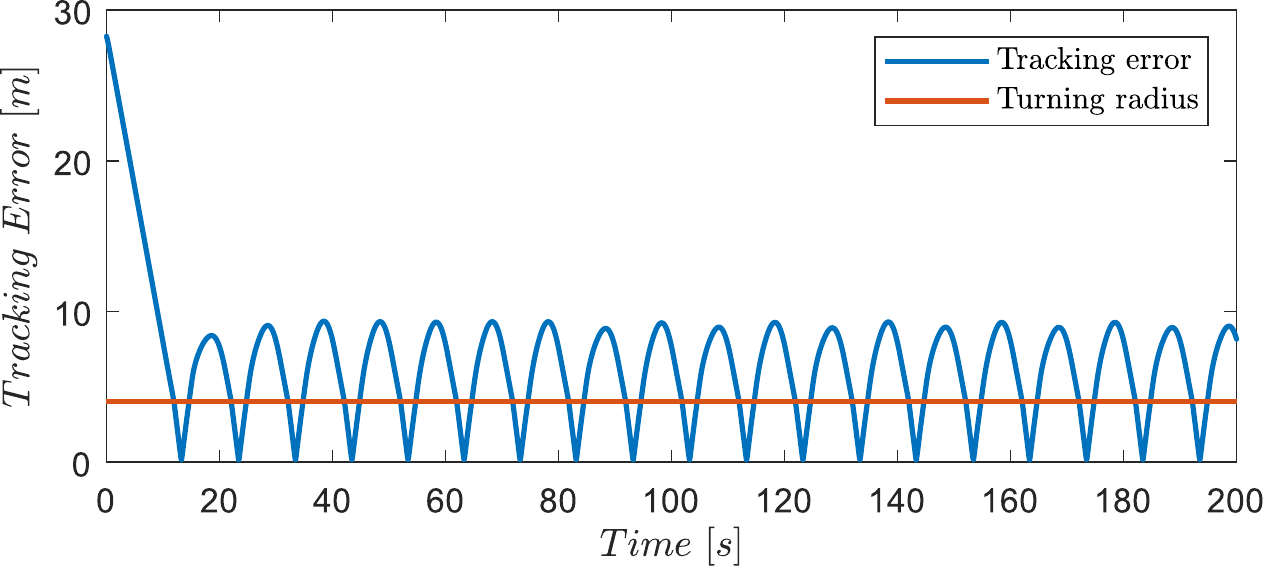}
	\vspace{-8pt}
	\captionof{figure}{Evolution of the tracking error for a single pursuer-evader system.}\label{error_pursuer}
\end{center}
\end{figure}

Figures~\ref{trajectory_pursuer} shows the trajectories of the pursuer and the evader, with the goal for the evader set to to point $(150,60)$. It can be observed that the evader moves towards the goal while the pursuer is far away and starts evasive maneuvers when it gets close to it, by entering its non-holonomic region. Figure~\ref{error_pursuer} displays the tracking error, defined as the distance between the pursuer and the evader, which is almost periodic. This is because the evader's maneuver forcing the pursuer to circle back. The peak tracking error after the pursuer catches up is slightly more than twice the turning radius, as expected.

\vspace{-1mm}
\subsection{Multiple Pursuers - Adversarial Evader}
While the previous section had only one pursuer, this simulation considers the case of two pursuers and a single evader. Having multiple pursuers means there must be cooperation between them in order to optimally utilize resources. Thus, a pursuer can no longer make decisions solely based on the position of the evader relative to itself. The positions of the rest of the pursuers must also be factored in.  Thus we redefine the expression for $g(x,u)$ to include these parameters as shown below for the case of two pursuers. Let $d_{\textrm{p}}$ be the distance between the two pursuers, and let
{\small\begin{align} \label{g_multi_pursuer}
g(x(t),u(t)) := \int_{t}^{t+T} &\bigg\{\beta _1 (d^2_1(\tau)+d^2_2(\tau)) + \beta _2 \frac{d^2_1(\tau)d^2_2(\tau)}{d^2_1(\tau)+d^2_2(\tau)} \nonumber \\ & \quad + \beta _3 e^{-\gamma d_\textrm{p}(\tau)}\bigg\}\textrm{d}\tau,\ \forall t\geq0.    
\end{align}}
The first term ensures that the pursuers remain close to the evader, while the second term encourages cooperation between agents. The last term is added to repel pursuers apart if they come close to each other, as having multiple pursuers in close vicinity of each other is sub-optimal. 

\begin{figure}[!ht]
\vspace{6pt}
\begin{center}
	\includegraphics[width=0.9\linewidth]{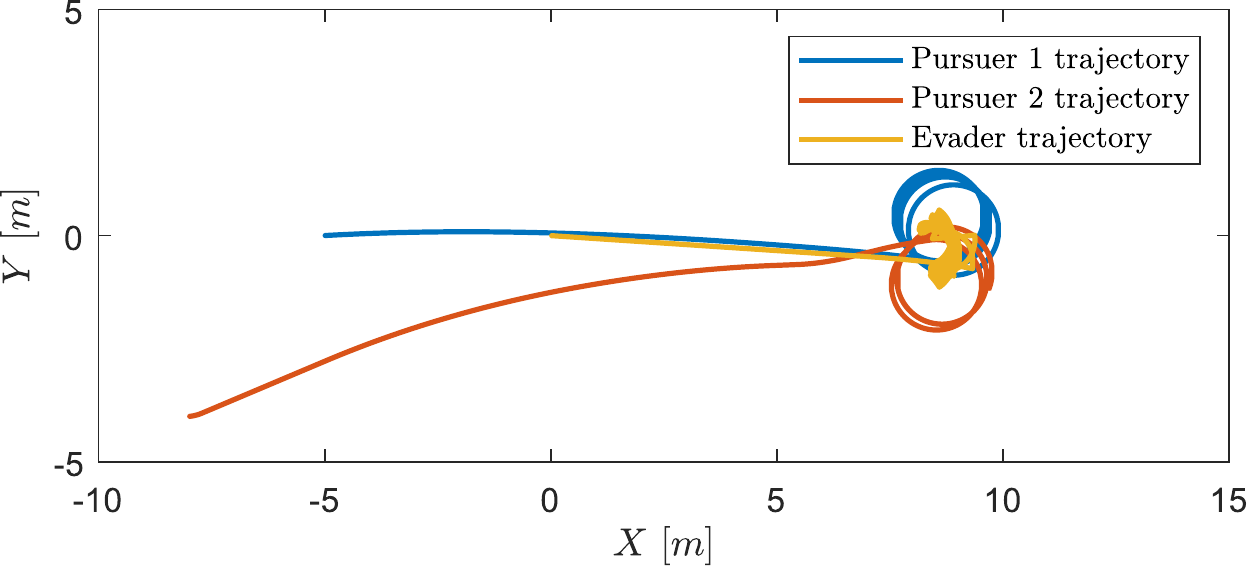}
	\vspace{-8pt}	
	\caption{ Trajectories for the two pursuer-single evader  system.}\label{trajectory_2pursuers}
\end{center}
\begin{center}
	\includegraphics[width=0.9\linewidth]{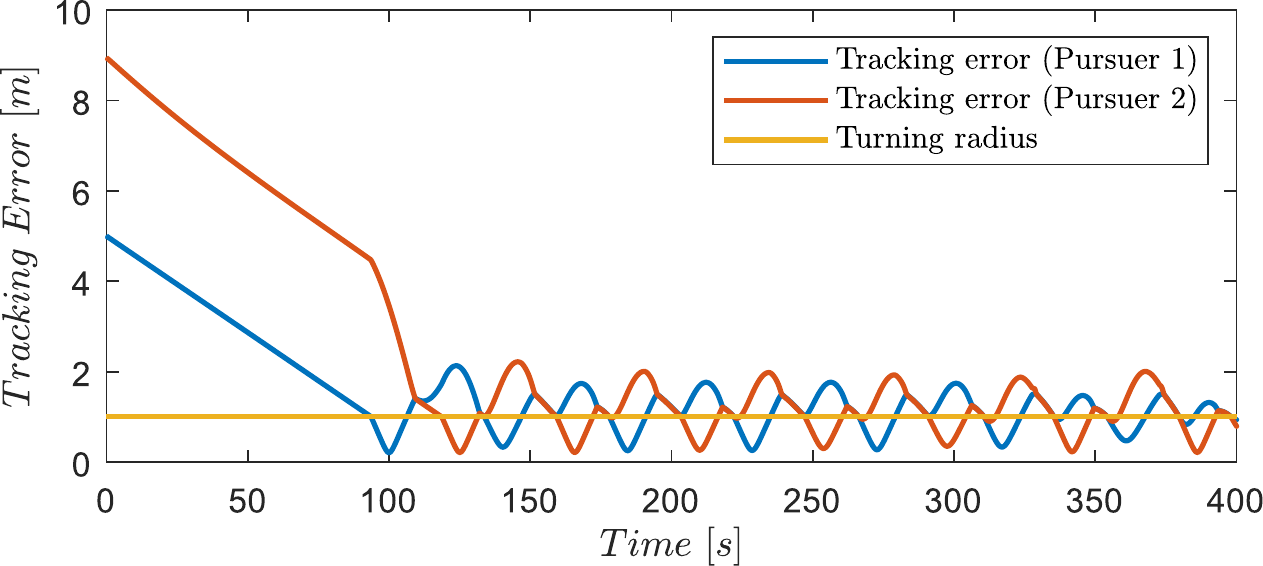}
	\vspace{-8pt}
	\caption{Evolution of the tracking error for the two pursuer-single evader system.}\label{error_2pursuer}
\end{center}
\end{figure}

Figure~\ref{trajectory_2pursuers} shows the trajectories of the pursuers and the evader when the goal for the evader is set to the point $(15,-1)$. In this case, the pursuers close in on the evader and trap it away from its goal due to their cooperative behavior. The evader is forced to continuously perform evasive maneuvers as the other pursuer closes in when the first has to make a turn. This can be seen more clearly in the tracking error plot given in Figure~\ref{error_2pursuer}. After catching up with the evader, it can be seen that when one pursuer is at its maximum distance, the other is at its minimum. 
The results achieved show good coordination between the  pursuers and low tracking error and are qualitatively comparable to \cite{quintero2016robust}. 

\begin{figure}[!ht]
\vspace{6pt}
\begin{center}
	\includegraphics[width=0.9\linewidth]{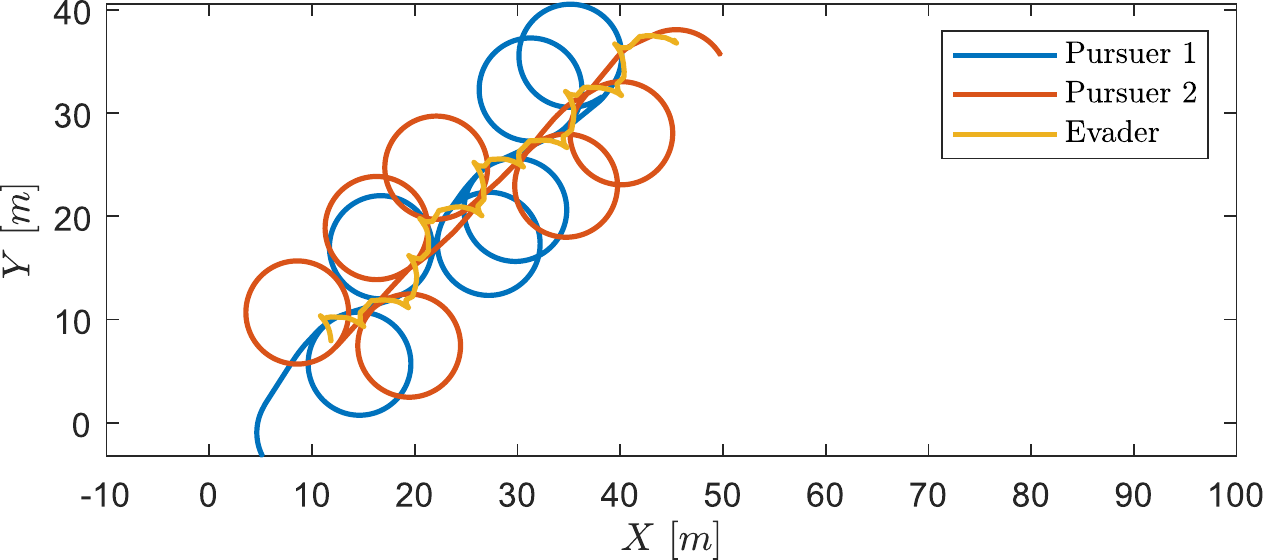}
	\vspace{-8pt}	
	\caption{Trajectories for two pursuers-single evader system with   learning.}\label{fig_nn_trajectory}
\end{center}

\begin{center}
	\includegraphics[width=0.9\linewidth]{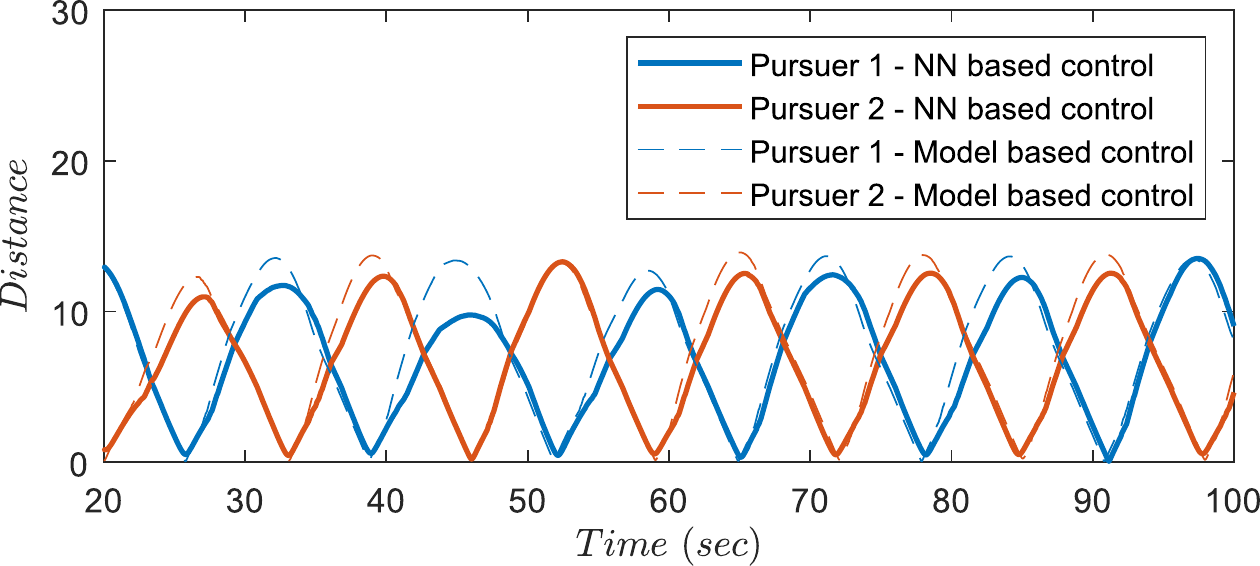}
	\vspace{-8pt}	
	\caption{ Evolution of the tracking error for the systems with and without learning.}\label{fig_nn_dist}
\end{center}
\begin{center}
	\includegraphics[width=0.9\linewidth]{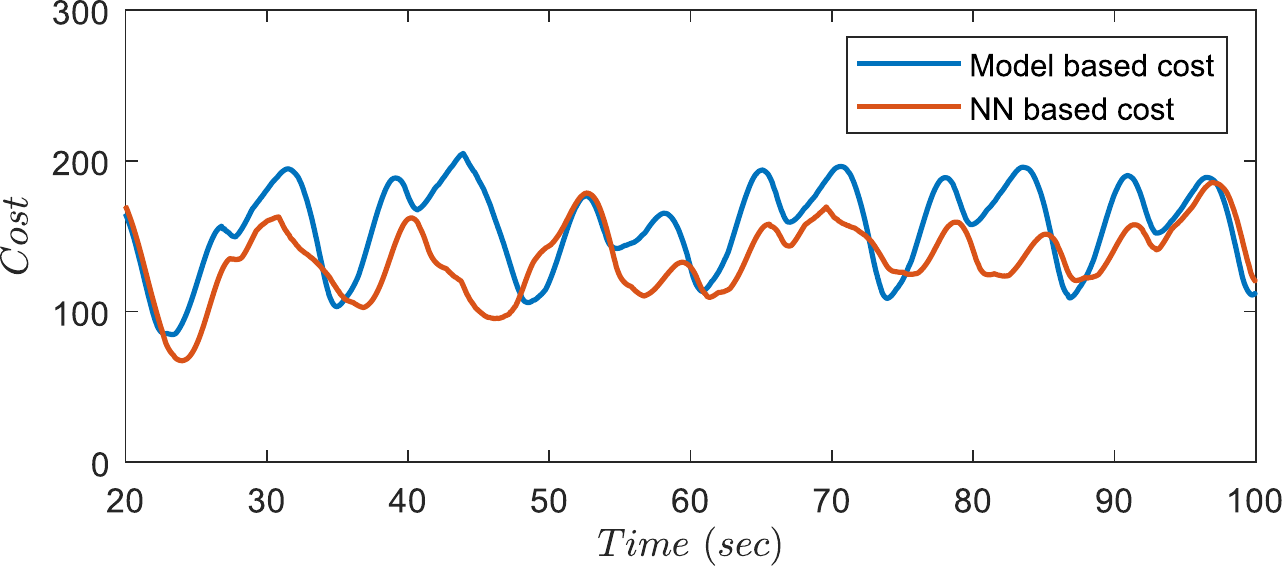}
	\vspace{-8pt}
	\captionof{figure}{{Total cost for the system with and without learning. }}\label{fig_nn_cost}
\end{center}
\end{figure}

Lastly,  we present the results under the learning-based prediction. In Figure~\ref{fig_nn_dist}, we present a comparative result of the tracking error of the model-based algorithm vis-\`a-vis the NN-based control. Figure~\ref{fig_nn_cost} showcases the quality of the performance of the proposed algorithm based on the game theoretic cost metric. From these figures, it can be seen that the NN structure offers fast predictive capabilities to the controller; hence the overall performance is comparable to the model based control.


\section{Conclusion and Future Work}
This work extends the framework of prediction-based nonlinear tracking in the context of pursuit evasion games.
We present results for vehicle pursuit of agnostic targets, modeled as moving along known trajectories, as well as adversarial target tracking, where the evader evolves according to game-theoretic principles. Furthermore, to obviate the need for explicit knowledge of the evader's strategy, we employ learning algorithms alongside the predictive controller.
The overall algorithm is shown to produce comparable results to those in the literature, while it precludes the need for solving an optimal control problem. 

Future work will focus on developing robustness guarantees will allow for more realistic scenarios, where noise and external disturbances are taken into consideration.

\balance

\bibliographystyle{IEEEtran}
\bibliography{learning_bib,Paper}

\begin{thebibliography}{10}
\providecommand{\url}[1]{#1}
\csname url@samestyle\endcsname
\providecommand{\newblock}{\relax}
\providecommand{\bibinfo}[2]{#2}
\providecommand{\BIBentrySTDinterwordspacing}{\spaceskip=0pt\relax}
\providecommand{\BIBentryALTinterwordstretchfactor}{4}
\providecommand{\BIBentryALTinterwordspacing}{\spaceskip=\fontdimen2\font plus
\BIBentryALTinterwordstretchfactor\fontdimen3\font minus
  \fontdimen4\font\relax}
\providecommand{\BIBforeignlanguage}[2]{{%
\expandafter\ifx\csname l@#1\endcsname\relax
\typeout{** WARNING: IEEEtran.bst: No hyphenation pattern has been}%
\typeout{** loaded for the language `#1'. Using the pattern for}%
\typeout{** the default language instead.}%
\else
\language=\csname l@#1\endcsname
\fi
#2}}
\providecommand{\BIBdecl}{\relax}
\BIBdecl

\bibitem{devasia1996nonlinear}
S.~Devasia, D.~Chen, and B.~Paden, ``Nonlinear inversion-based output
  tracking,'' \emph{IEEE Transactions on Automatic Control}, vol.~41, no.~7,
  pp. 930--942, 1996.

\bibitem{martin1996different}
P.~Martin, S.~Devasia, and B.~Paden, ``A different look at output tracking:
  control of a vtol aircraft,'' \emph{Automatica}, vol.~32, no.~1, pp.
  101--107, 1996.

\bibitem{Isidori90}
A.~Isidori and C.~Byrnes, ``Output regulation of nonlinear systems,''
  \emph{IEEE Trans. Automat. Control}, vol.~35, pp. 131--140, 1990.

\bibitem{Khalil98}
H.~Khalil, ``On the design of robust servomechanisms for minimum phase
  nonlinear systems,'' \emph{Proc. 37th IEEE Conf. Decision and Control, {\rm
  Tampa, FL}}, pp. 3075--3080, 1998.

\bibitem{allgower2012nonlinear}
F.~Allg{\"o}wer and A.~Zheng, \emph{Nonlinear model predictive control}.\hskip
  1em plus 0.5em minus 0.4em\relax Birkh{\"a}user, 2012, vol.~26.

\bibitem{Rawlings17}
J.~Rawlings, D.~Mayne, and M.~Diehl, \emph{Model Predictive Control: Theory,
  Computation, and Design, 2nd Edition}.\hskip 1em plus 0.5em minus 0.4em\relax
  Nob Hill, LLC, 2017.

\bibitem{Wardi17}
Y.~Wardi, C.~Seatzu, M.~Egerstedt, and I.~Buckley, ``Performance regulation and
  tracking via lookahead simulation: Preliminary results and validation,'' in
  \emph{56th IEEE Conf. on Decision and Control, {\rm Melbourne, Australia,
  December 12-15}}, 2017.

\bibitem{Wardi18}
Y.~Wardi, C.~Seatzu, and M.~Egerstedt, ``Tracking control via variable-gain
  integrator and lookahead simulation: Application to leader-follower
  multiagent networks,'' in \emph{Sixth IFAC Conference on Analysis and Design
  of Hybrid Systems (2018 ADHS)l, {\rm Oxford, the UK, July 11-13}}, 2018.

\bibitem{shivam2018tracking}
S.~Shivam, I.~Buckley, Y.~Wardi, C.~Seatzu, and M.~Egerstedt, ``Tracking
  control by the newton-raphson flow: Applications to autonomous vehicles,'' in
  \emph{European Control Conference, {\rm Naples, Italy, June 25-28}}, 2018.

\bibitem{saridis1983intelligent}
G.~Saridis, ``Intelligent robotic control,'' \emph{IEEE Transactions on
  Automatic Control}, vol.~28, no.~5, pp. 547--557, 1983.

\bibitem{haykin2009neural}
S.~S. Haykin, \emph{Neural networks and learning machines}.\hskip 1em plus
  0.5em minus 0.4em\relax Pearson Upper Saddle River, 2009, vol.~3.

\bibitem{vrabie2013optimal}
D.~Vrabie, K.~G. Vamvoudakis, and F.~L. Lewis, \emph{Optimal adaptive control
  and differential games by reinforcement learning principles}.\hskip 1em plus
  0.5em minus 0.4em\relax IET, 2013, vol.~2.

\bibitem{narendra1990identification}
K.~S. Narendra and K.~Parthasarathy, ``Identification and control of dynamical
  systems using neural networks,'' \emph{IEEE Transactions on neural networks},
  vol.~1, no.~1, pp. 4--27, 1990.

\bibitem{vamvoudakis2010online}
K.~G. Vamvoudakis and F.~L. Lewis, ``Online actor--critic algorithm to solve
  the continuous-time infinite horizon optimal control problem,''
  \emph{Automatica}, vol.~46, no.~5, pp. 878--888, 2010.

\bibitem{bishop1995neural}
C.~M. Bishop \emph{et~al.}, \emph{Neural networks for pattern
  recognition}.\hskip 1em plus 0.5em minus 0.4em\relax Oxford university press,
  1995.

\bibitem{bhasin2013novel}
S.~Bhasin, R.~Kamalapurkar, M.~Johnson, K.~G. Vamvoudakis, F.~L. Lewis, and
  W.~E. Dixon, ``A novel actor--critic--identifier architecture for approximate
  optimal control of uncertain nonlinear systems,'' \emph{Automatica}, vol.~49,
  no.~1, pp. 82--92, 2013.

\bibitem{vamvoudakis2017q}
K.~G. Vamvoudakis, ``Q-learning for continuous-time linear systems: A
  model-free infinite horizon optimal control approach,'' \emph{Systems \&
  Control Letters}, vol. 100, pp. 14--20, 2017.

\bibitem{weber2009data}
B.~G. Weber and M.~Mateas, ``A data mining approach to strategy prediction,''
  in \emph{2009 IEEE Symposium on Computational Intelligence and Games}.\hskip
  1em plus 0.5em minus 0.4em\relax IEEE, 2009, pp. 140--147.

\bibitem{alpcan2010network}
T.~Alpcan and T.~Ba{\c{s}}ar, \emph{Network security: A decision and
  game-theoretic approach}.\hskip 1em plus 0.5em minus 0.4em\relax Cambridge
  University Press, 2010.

\bibitem{pesch1995synthesis}
H.~J. Pesch, I.~Gabler, S.~Miesbach, and M.~H. Breitner, ``Synthesis of optimal
  strategies for differential games by neural networks,'' in \emph{New Trends
  in Dynamic Games and Applications}.\hskip 1em plus 0.5em minus 0.4em\relax
  Springer, 1995, pp. 111--141.

\bibitem{Kanellopoulos19}
A.~Kanellopoulos, K.~Vamvoudakis, and Y.~Wardi, ``Predictive learning via
  lookahead simulation,'' in \emph{AIAA Scitech 2019 Forum, {\rm San Diego,
  California, January 7-11}}, 2019.

\bibitem{quintero2016robust}
S.~A. Quintero, D.~A. Copp, and J.~P. Hespanha, ``Robust uav coordination for
  target tracking using output-feedback model predictive control with moving
  horizon estimation,'' in \emph{American Control Conference, {\rm Chicago,
  Illinois, July 1-3}}, 2015.

\bibitem{lavalle2006planning}
S.~M. LaValle, \emph{Planning algorithms}.\hskip 1em plus 0.5em minus
  0.4em\relax Cambridge university press, 2006.

\bibitem{isaacs1999differential}
R.~Isaacs, \emph{Differential games: a mathematical theory with applications to
  warfare and pursuit, control and optimization}.\hskip 1em plus 0.5em minus
  0.4em\relax Courier Corporation, 1999.

\bibitem{basar1999dynamic}
T.~Basar and G.~J. Olsder, \emph{Dynamic noncooperative game theory}.\hskip 1em
  plus 0.5em minus 0.4em\relax Siam, 1999, vol.~23.

\bibitem{lewis1998neural}
F.~Lewis, S.~Jagannathan, and A.~Yesildirak, \emph{Neural network control of
  robot manipulators and non-linear systems}.\hskip 1em plus 0.5em minus
  0.4em\relax CRC Press, 1998.

\end{thebibliography}

\end{document}